\documentclass[11pt,a4paper]{article}
\usepackage{jheppub}
\title{A New Model of Holographic Dark Energy with Action Principle}

\author[a]{Miao Li,}

\author[b]{Rong-Xin Miao}

\affiliation[a]{Kavli Institute for Theoretical Physics, State Key
Laboratory of Theoretical Physics, Institute of
Theoretical Physics, Chinese Academy of Sciences, \\
Beijing 100190, People's Republic of China.\footnote{mli@itp.ac.cn}
}

\affiliation[b]{Interdisciplinary Center for Theoretical Study, University of Science and Technology of China,\\
Hefei, Anhui 230026, People's Republic of
China.\footnote{mrx11@mail.ustc.edu.cn}}

\abstract{We propose a new model of holographic dark energy with an
action. It is the first time that one can derive a HDE model from
the action principle. The puzzles of causality and circular logic
about HDE have been completely solved in this model. The evolution
of the universe only depends on the present state of the universe,
clearly showing that it obeys the law of causality. Furthermore, the
use of future event horizon as a present cut-off is not an input but
automatically follows from equations of motion. Interestingly, this
new model is very similar to the initial one of Li except a new term
which may be explained as dark radiation.}

\begin{document}

\maketitle
\section{Introduction}
The dark energy problem is a longstanding problem ever since the
discovery of the accelerating expansion of the universe \cite{Riess}.
For a recent review of dark energy, please refer to \cite{Li0}.
Based on an idea of Cohen et al \cite{Cohen,Horava}, Li proposed the
first model of the holographic dark energy (HDE) which can drive the
accelerating expansion \cite{Li1}. Though it is in good agreement
with observational data \cite{Xiaodong}, it causes some criticisms
due to its use of the future event horizon as a present cut-off.
These criticisms can be summarized as the causality problem and the
circular logic problem. The causality problem: it seems that the
evolution of the universe depends on the future information about the universe,
the future event horizon. Beside, the equations of motion are
non-local since the future event horizon is defined globally. The
circular logic problem: the future event horizon exists only in an
accelerating universe. How can one use an assumption based on the
accelerating expansion to explain the accelerating expansion? For
recent interesting discussions of these problems, please see
\cite{Kim}. For proposals using other infrared cut-off, see
\cite{age}-\cite{wu}.

In this paper, we solve the causality problem and circular logic
problem of HDE completely. First, we derive a new HDE model from the
action principle. We find that we can rewrite all equations of
motion (both the new and the initial HDE models) in local forms.
Then the evolution of the universe only depends on the present
initial conditions, which clearly obeys the law of causality. What
is more, we do not need  to assume the future event horizon as the
infrared cut-off but only a local equation $\dot{L}=-\frac{1}{a}$.
Magically, the equations of motion force the cut-off to be exactly
the future event horizon $aL$. So it is not that the present cut-off
depends on the future event horizon but conversely the future event
horizon is completely determined by present cut-off through
equations of motion.

This paper is arranged as follows. In Sec.~2, we derive a new HDE
model from the action principle. We find exact solutions with matter
and radiation and investigate cosmology in this model. In Sec.~3, we
study models with other infrared cut-off in the action. We conclude
in Sec.~4.

\section{New HDE model from action principle}
In this section, we derive a new model of holographic dark energy
from the action principle. We find exact solutions of this model and
prove that the cut-off is exactly the future event horizon. We also
investigate cosmology of this model and find that radiation is
dominant in early time, while dark energy is dominant in late time.

\subsection{The general theory}

Consider the Robertson-Walker metric
\begin{eqnarray}\label{metric}
ds^2=-N^2(t)dt^2+a^2(t)[\frac{dr^2}{1-\kappa r^2}+r^2d\Omega^2]
\end{eqnarray}
and the following action
\begin{eqnarray}\label{action}
S=\frac{1}{16\pi}\int dt[\sqrt{-g}(R-\frac{2c}{a^2(t)
L^2(t)})-\lambda(t)(\dot{L}(t)+\frac{N(t)}{a(t)})]+S_M ,
\end{eqnarray}
where $R$ is the Ricci scalar, $\sqrt{-g}=N a^3$ (we have integrated
the $r, \theta, \phi$ parts) and $S_M$ denotes the action of all
matter fields (we use $M$ to denote all matter fields, $m$ to denote
matter without pressure and $r$ to denote radiation). In following
derivations, we first take the variations of $N, a, \lambda, L$ and
then redefine $Ndt$ as $dt$. We obtain
\begin{eqnarray}\label{equation0}
(\frac{\dot{a}}{a})^2+\frac{\kappa}{a^2}=\frac{c}{3a^2L^2}+\frac{\lambda}{6a^4}+\frac{8\pi}{3}\rho_M,\nonumber \\
\frac{2\ddot{a}a+\dot{a}^2+\kappa}{a^2}=\frac{c}{3a^2L^2}-\frac{\lambda}{6a^4}-8\pi
p_M,
\end{eqnarray}
and
\begin{eqnarray}\label{equation1}
\dot{L}&=&-\frac{1}{a},\ \ \ \ \ L=\int_t^{\infty}\frac{dt'}{a(t')}+L(\infty)\nonumber \\
\dot{\lambda}&=&-\frac{4ac}{L^3},\ \
\lambda=-\int_0^{t}dt'\frac{4a(t')c}{L^3(t')}+\lambda(0).
\end{eqnarray}
We want to mention that we shall prove $L(\infty)=0$ using the
asymptotic solutions to be derived below. So, quite interestingly,
$aL$ is exactly the future event horizon. Besides, from
eq.(\ref{equation0}), it is easy to see that the $\lambda(0)$
term behaves the same way as radiation, thus can be a candidate for dark
radiation \cite{darkradiation}. For the purpose of solving equations
of motion, we can always redefine $\lambda(t)$ as
$\lambda(t)-\lambda(0)$, $\rho_r$ as $\rho_r
+\frac{\lambda(0)}{16\pi a^4}$, to let $\lambda(0)=0$. From
eq.(\ref{equation0}), we can derive
\begin{eqnarray}\label{equation3}
\frac{\ddot{a}}{a}=-\frac{\lambda}{6a^4}-\frac{4\pi}{3}(\rho_M+3p_M),
\end{eqnarray}
Note that we have $\lambda<0$ for large enough time from
eq.(\ref{equation1}) ($c>0$). We can ignore the matter effects
for large enough time when dark energy is dominant. So
eq.(\ref{equation3}) implies the accelerating expansion of the universe.

Let us now try to find out solutions of the above equations.
Fortunately, we can solve eqs.(\ref{equation0}-\ref{equation3})
exactly with general $c, \kappa$, $\rho_m\propto \frac{1}{a^3}$ and
$\rho_r\propto \frac{1}{a^4}$. As a simple example, it is easy to
observe that there is a de Sitter solution with $(a=e^{Ht},
L=e^{-Ht}/H, \lambda=-6H^2e^{-4Ht})$ and $(c=6, \kappa=0,
\rho_M=p_M=0)$. We postpone discussing the general case to the next
section.

Let us go on to discuss the asymptotic solutions for large
enough time. Using eq.(\ref{equation0}), we can derive
\begin{eqnarray}\label{equation4}
\frac{d(\dot{a}a)
}{dt}=\frac{c}{3L^2}-\kappa+\frac{4\pi}{3}a^2(\rho_M-3p_M).
\end{eqnarray}
Note that $\frac{c}{3L^2}$ is the only increasing function of time (
$L$ is a decreasing function) on the right hand of
eq.(\ref{equation4}), which becomes dominant for large enough time.
So, we have
\begin{eqnarray}\label{equation5}
\frac{d(\dot{a}a) }{dt}=\frac{c}{3L^2}
\end{eqnarray}
asymptotically. Applying $\dot{L}=-\frac{1}{a}$, we can rewrite the above equation as
\begin{eqnarray}\label{equation5}
\frac{d^2 a }{dL^2}=\frac{c a}{3L^2},
\end{eqnarray}
with the general solution
\begin{eqnarray}\label{solution}
a=c_1 L^{\frac{1-\sqrt{1+\frac{4}{3}c}}{2}}+c_2
L^{\frac{1+\sqrt{1+\frac{4}{3}c}}{2}}\sim  L^{\frac{1-\sqrt{1+\frac{4}{3}c}}{2}}.
\end{eqnarray}
Using $\dot{L}=-\frac{1}{a}$ and the asymptotic solution $a\sim  L^{\frac{1-\sqrt{1+\frac{4}{3}c}}{2}}$, we can derive
\begin{eqnarray}
&a&\sim t^{\frac{1-\sqrt{1+\frac{4}{3}c}}{3-\sqrt{1+\frac{4}{3}c}}},\ \ \ \ \ \ \ \ \ \ \ c>6\label{solution1} \\
&a&\sim e^{c_3 t}, c_3>0, \ \ \ \ \ \ \ \ \ c=6\label{yessolution1} \\
&a&\sim
(c_4-t)^{\frac{1-\sqrt{1+\frac{4}{3}c}}{3-\sqrt{1+\frac{4}{3}c}}},\
\ c<6.\label{bigripsolution}
\end{eqnarray}
Eq.(\ref{bigripsolution}) implies that time would end at $t=c_4$. Using
eq.(\ref{solution}) with $c<6$, we can derive $aL\rightarrow 0$ when
$L\rightarrow 0$ or equivalently $t\rightarrow c_4$. Thus the future
event horizon shrinks to a point ($aL\rightarrow 0$) at $t=c_4$.
Then time would end. In fact, below we shall see that we have $w<-1$
when $c<6$. So there would be a big rip for the universe with $c<6$.

From the state parameter of dark energy
\begin{eqnarray}\label{w}
w\equiv\frac{p_{de}}{\rho_{de}}=\frac{\lambda L^2-2ca^2}{3\lambda
L^2+6ca^2}=\frac{-3+2c+\sqrt{9+12c}}{3(-3-2c+\sqrt{9+12c})},
\end{eqnarray}
we get
\begin{eqnarray}\label{Aw}
w=-1,\ \  c=6,\nonumber\\
-1< w <-\frac{1}{3}, \ \ c> 6,\nonumber\\
w <-1,\ \ \  c<6.
\end{eqnarray}
Note that
\begin{eqnarray}\label{Denergy}
\rho_{de}\propto a^{-3(1+w)}>\rho_M,
\end{eqnarray}
consistent with our above assumption that dark energy is
dominant for large enough time.

The asymptotic solutions eqs.(\ref{solution1},\ref{yessolution1})
imply $a(\infty)\rightarrow\infty$ and eq.(\ref{bigripsolution})
implies $a(t=c_4)\rightarrow\infty$. So we have
$L(\infty)\rightarrow 0$ or $L(t=c_4)\rightarrow 0$ from
eq.(\ref{solution}). Thus, it is very interesting that $aL$ is
exactly the future event horizon. It should be mentioned that the
above method can also apply to the initial HDE model. Using
asymptotic solutions, we need not to assume that $aL$ is exactly the
future event horizon but only $\dot{L}=-\frac{1}{a}$. Then all the
equations of the initial HDE model become local and depend on only
the present initial conditions.

Let us comment on our results. First, it is the first model of
holographic dark energy with action principle. Second, it implies
the accelerating expansion of the universe. Third,
eqs.(\ref{equation0},\ref{equation1}) show that the evolution of the
universe only depends on the present initial conditions $a, \dot{a},
L, \lambda$. So, it is clear that our HDE model obeys the law of
causality. Furthermore, the use of the future event horizon as a
cut-off is not an input in our model. Instead, the cut-off $aL$
turns out to be the future event horizon automatically from
equations of motion. So the long-standing problem of HDE  ``Why does
the present evolution of the universe depend on the future of
universe" has been solved. In fact, the evolution of the universe
only depends on present conditions but equations of motion force the
future event horizon to be the present cut-off $aL$ magically. In a
word, in this model we answer the above question by ``It is natural
that the future of the universe (future event horizon) depends on
the present conditions ($aL$) of the universe." Fourth, this model
includes a de Sitter solution. Fifth, the energy density of dark
energy
\begin{eqnarray}\label{energy}
\rho_{de}=\frac{1}{8\pi}(\frac{c}{a^2L^2}+\frac{\lambda}{2a^4})
\end{eqnarray}
will remain positive if it is positive at the beginning and
$\dot{a}>0$. Finally, from eq.(\ref{action}) it is interesting to
note that $(-\lambda)$ is the conjugate momentum of the cut-off $L$.
Or equivalently, $\lambda$ is the energy associated with the
conformal time $\eta=-L$.
\subsection{The exact solutions}
In this section, we shall solve equations
eqs.(\ref{equation0}-\ref{equation3}) exactly with matter ($\rho_m=
\frac{3b}{4\pi a^3}, p_m=0)$ and radiation $(\rho_r=\frac{3d}{8\pi
a^4}, p_r=\frac{1}{3}\rho_r)$. Thus, we have $\rho_M=\frac{3b}{4\pi
a^3}+\frac{3d}{8\pi a^4}, p_M=\frac{d}{8\pi a^4}$. For simplicity,
we focus on the case with $\kappa=0$. It should be stressed that one
can solve these equations exactly with general $\kappa$.

Redefine a new time $dL=-d\eta=-\frac{dt}{a}$, we can rewrite
eq.(\ref{equation4}) as
\begin{eqnarray}\label{Appequation}
\frac{d^2 a }{dL^2}=\frac{c a}{3L^2}+b.
\end{eqnarray}
Note that the above equation is independent of the parameter $d$. It
seems that radiation does not affect the evolution of the scale
factor $a$. This is however not the case. As we have mentioned
before, we can always redefine $\lambda$ as $(\lambda-\lambda(0))$ ,
$d$ as $(d+\frac{\lambda(0)}{6})$, to let $\lambda(0)=0$ (note
$\lambda(0)$ means $\lambda(t=0)$ instead of $\lambda(L=0)$). Then,
using $\lambda(0)=0$ we can relate the radiation parameter $d$ to
the scale factor $a$.

 The general solution of eq.(\ref{Appequation}) is
\begin{eqnarray}
&a&=c_1\frac{1}{L}+c_2 L^2+\frac{1}{3}b L^2\ln L,\ \ \ \ \ \ \ \ \ \ \ \ \ \ \ \ \ \ \  \ \ \ c=6,\label{Appsolution}\\
&a&=c_1 L^{\frac{1-\sqrt{1+\frac{4c}{3}}}{2}}+c_2
L^{\frac{1+\sqrt{1+\frac{4c}{3}}}{2}}-\frac{3b}{c-6}L^2, \ \ \ \
c\neq 6.\label{Appsolution1}
\end{eqnarray}

Let us first discuss the case with $c=6$. Using $dt=-a dL$, we get
\begin{eqnarray}\label{Apptime}
t=-c_1\ln L-\frac{1}{3}c_2 L^3-\frac{1}{9}bL^3\ln L
+\frac{b}{27}L^3.
\end{eqnarray}
From eq.(\ref{equation3}), we can derive
\begin{eqnarray}\label{Applambda}
&&\lambda=-6[a^3\ddot{a}+a b+d]=-6[a a''-a'^2 +a b+d]\nonumber\\
&&=-6d+\frac{2 b^2 L^2}{3} - 4 b L^2 c_2 + 12 L^2 c_2^2 - \frac{16 b c_1}{L} - \frac{48c_1c_2}{L} - \frac{6 c_1^2}{L^4}\nonumber\\
&&\ \ \ - \frac{4}{3} b^2 L^2 \ln L+8 b L^2 c_2 \ln L - \frac{16
bc_1\ln L}{L} + \frac{4}{3} b^2 L^2 (\ln L)^2,
\end{eqnarray}
with $a'=\frac{d a}{dL}$. From the above equation, we can derive
$\lambda'$ while we can also derive $\lambda'=\frac{4a^2 c}{L^3}$
from eq.(\ref{equation1}). One can check that these two results are
equal to each other, which implies that we have got the
self-consistent solutions to eqs.(\ref{equation0}-\ref{equation3}).
With eqs.(\ref{Appsolution},\ref{Applambda}), we can obtain the
asymptotic state parameter as
\begin{eqnarray}\label{Adew}
&w&\rightarrow -1, \ \ \ \ \ L\rightarrow 0,\nonumber\\
&w&\rightarrow -\frac{1}{3}, \ \ \ \ L\rightarrow L_0,
\end{eqnarray}
where $L_0$ is defined by $a(L_0)=0$, denoting the beginning
time of the universe $t=0$.

Now let us turn to discussing the case with $c\neq6$ briefly.  With
eq.(\ref{Appsolution1}), we can derive
\begin{eqnarray}\label{Applambda1}
\lambda=-6[a^3\ddot{a}+a b+d]=-6[a a''-a'^2 +a b+d]=-6d+...
\end{eqnarray}
One can check that eqs.(\ref{Appsolution1},\ref{Applambda1}) satisfy
all of eqs.(\ref{equation0}-\ref{equation3}). From
eq.(\ref{equation1}), it is easy to observe that $\lambda(0)=\lim
\lambda|_{c\rightarrow0}$. Applying the condition
\begin{eqnarray}\label{limit}
\lim \lambda|_{c\rightarrow0}=-6(2bc_1+d-c_2^2)=0,
\end{eqnarray}
we can derive
\begin{eqnarray}\label{condition}
c_2^2=d+2c_1 b,
\end{eqnarray}
which shows that radiation does affect the evolution of the scale
factor $a$ eq.(\ref{Appsolution1}).

One can derive the asymptotic state parameter as
\begin{eqnarray}\label{Adew1}
&&-1< w\rightarrow\frac{-3+2c+\sqrt{9+12c}}{3(-3-2c+\sqrt{9+12c})} <-\frac{1}{3}, \ \ \ L\rightarrow 0,\nonumber\\
&& w\rightarrow -\frac{1}{3}, \ \ \ \ \ \ \ \ \ \ \ \ \ \ \ \ \ \ \
\ \ \ \ \ \ \ \ \ \ \ \ \ \ \ \ \ \ \ \ \ \ \ \ \ \ L\rightarrow
L_0,
\end{eqnarray}
for $c>6$ and similarly
\begin{eqnarray}\label{Adew2}
&&w\rightarrow\frac{-3+2c+\sqrt{9+12c}}{3(-3-2c+\sqrt{9+12c})} <-1,\ \  \ \ \ \ \ \ \ \ \ \ L\rightarrow 0,\nonumber\\
&&w\rightarrow -\frac{1}{3}, \ \ \ \ \ \ \ \ \ \ \ \ \ \ \ \ \ \ \ \
\ \ \ \ \ \ \ \ \ \ \ \ \ \ \ \ \ \  \ \ \ \ \ \ \ \ L\rightarrow
L_0
\end{eqnarray}
for $0<c<6$.

\subsection{Cosmology}

In this section, we briefly discuss cosmology in our model. It is
necessary to check that radiation is dominant in early time and
$a\sim \sqrt{t}$ thus consistent with the standard
 cosmology. In early time, we have
\begin{eqnarray}\label{earlya}
a(L_0)=0,\ \ \ a(L)\approx a'(L_0)(L-L_0),\ \ a'(L_0)<0.
\end{eqnarray}
Applying $\dot{L}=-\frac{1}{a}$, we get
\begin{eqnarray}\label{earlyt}
t\approx-\frac{a'(L_0)}{2}(L-L_0)^2.
\end{eqnarray}
Thus, we obtain the expected result
\begin{eqnarray}\label{earlyt}
a(t)\approx\sqrt{-2a'(L_0) t}\ ,
\end{eqnarray}
implying that radiation is dominant in early time. Using the
above solutions and $\lambda(0)=0$, we can derive
\begin{eqnarray}\label{earlydarkenergy}
w_{de}\approx -\frac{1}{3},\ \ \Omega_{de}\approx 0,\ \
\Omega_{r}\approx 1,
\end{eqnarray}
which is consistent with eqs.(\ref{Adew},\ref{Adew1},\ref{Adew2}) in
the early time limit $L\rightarrow L_0$. Using the exact solutions
eqs.(\ref{Appsolution},\ref{Appsolution1}), we can obtain the same
results as above. So our model has passed the check that radiation
is dominant in early time and $a\sim \sqrt{t}$. Thus, it would not
ruin standard results such as nuclear genesis.

Applying the asymptotic or exact solutions, we can easily find that
dark energy is dominant in the late time and state parameter behaves
as
\begin{eqnarray}\label{AwAw}
w=-1,\ \  c=6,\nonumber\\
-1< w <-\frac{1}{3}, \ \ c> 6,\nonumber\\
w <-1,\ \ \  c<6.
\end{eqnarray}
When $c=6$, the universe will approach de Sitter space asymptotically.
While for $c<6$, dark energy will behave as phantom in late time and
end up with a big rip. So it is necessary to estimate the value of
$c$.

Since the universe has turned to the phase of accelerating expansion
just recently, now we have $\frac{\ddot{a}}{a}\approx 0$ which leads
to $\frac{\lambda}{6a^4}\approx-\frac{4\pi}{3}\rho_m$ from
eq.(\ref{equation3}). Using $aL\approx \frac{1}{H}$ together with
$\rho_{de}\approx \frac{7}{3}\rho_m$, we have $\frac{c}{3
a^2L^2}\approx\frac{cH^2}{3}\approx\frac{17}{20}H^2$. It follows
that $c\approx 2.6$ which implies that, similar to the initial HDE
model, dark energy will behave as phantom in late time. It is
interesting to check our model with the observational data. We leave
it to the future works.

Finally, we want to talk about the radiation-like term $\lambda(0)$
in our model. We can separate the energy density
$(\frac{\lambda}{16\pi a^4}+\frac{c}{8\pi a^2L^2})$ and pressure
$(\frac{\lambda}{48\pi a^4}-\frac{c}{24\pi a^2L^2})$ into
eq.(\ref{equation0}) into the dark energy part and dark radiation
part:
\begin{eqnarray}\label{darkdiation}
&&\rho_{de}=\frac{\lambda-\lambda(0)}{16\pi a^4}+\frac{c}{8\pi a^2L^2},\nonumber\\
&&p_{de}=\frac{\lambda-\lambda(0)}{48\pi a^4}-\frac{c}{24\pi a^2L^2},\nonumber\\
&&\rho_{dr}=\frac{\lambda(0)}{16\pi a^4},\ \ \ \ \
p_{dr}=\frac{1}{3}\rho_{dr}.
\end{eqnarray}
By dark radiation, we mean that it has no interactions with other
fields. It originates from the initial condition of $\lambda$ and never
decays. Interestingly, observational evidences support the existence
of dark radiation \cite{darkradiation}. We shall study this topic in
details in the future work.

\section{Models with other cutoff}
In this section, we use the particle horizon and the Hubble horizon as the
cut-off in the action. We find that, similar to the initial HDE
model, neither of them can drive the accelerating expansion of the
universe.

\subsection{The particle horizon as the cutoff}

In this subsection, we shall use the particle horizon as a cut-off.
For simplicity, we focus on the case $\kappa=\rho_M=p_M=0$. We find
that this model has no accelerating solutions. Let us start with the
action
\begin{eqnarray}\label{action1}
S=\frac{1}{16\pi}\int dt[\sqrt{-g}(R-\frac{2c}{a^2(t)
L^2(t)})-\lambda(t)(\dot{L}(t)-\frac{N(t)}{a(t)})].
\end{eqnarray}
Following the same procedure of the above section, we obtain
\begin{eqnarray}\label{equation00}
&&(\frac{\dot{a}}{a})^2=\frac{c}{3a^2L^2}-\frac{\lambda}{6a^4},\nonumber \\
&&\frac{2\ddot{a}a+\dot{a}^2}{a^2}=\frac{c}{3a^2L^2},
\end{eqnarray}
and
\begin{eqnarray}\label{equation10}
\dot{L}&=&\frac{1}{a},\ \ \ \ \ L=\int_0^{t}\frac{dt'}{a(t')}+L(0)\nonumber \\
\dot{\lambda}&=&-\frac{4ac}{L^3},\ \
\lambda=-\int_0^{t}dt'\frac{4a(t')c}{L^3(t')}+\lambda(0).
\end{eqnarray}
From eq.(\ref{equation00}), we can derive
\begin{eqnarray}\label{equation30}
\frac{\ddot{a}}{a}=\frac{\lambda}{6a^4}.
\end{eqnarray}

If $c>0$, eq.(\ref{equation10}) implies $\lambda<0$ so that
$\frac{\ddot{a}}{a}<0$ for large enough time. On the other hand, eq.(\ref{equation00}) implies $\lambda<0$ if $c<0$. So this model
has no accelerating solutions in either case.

\subsection{The Hubble horizon as the cutoff}

Now let us turn to the case of the Hubble horizon as the cut-off. We
focus on the case $\kappa=0$. We begin with the following action
\begin{eqnarray}\label{action2}
S=\frac{1}{16\pi}\int dt[\sqrt{-g}(R-\frac{2c}{
L^2(t)})-\lambda(t)(L(t)-\frac{N(t)}{H})]+S_M.
\end{eqnarray}
Following a similar procedure as the above sections, we obtain
\begin{eqnarray}
&&(\frac{\dot{a}}{a})^2=\frac{c}{3L^2}-\frac{\lambda}{6a^2\dot{a}}+\frac{8\pi}{3}\rho_M=-\frac{c}{3}(\frac{\dot{a}}{a})^2+\frac{8\pi}{3}\rho_M,\label{equation002} \\
&&\frac{2\ddot{a}a+\dot{a}^2}{a^2}=\frac{c}{L^2}-\frac{\dot{\lambda}}{6a\dot{a}^2}-\frac{\lambda(\dot{a}^2-\ddot{a}a)}{3a^2\dot{a}^3}-8\pi p_M\nonumber \\
&&\ \ \ \ \ \ \ \ \ \ \ \ \ =
-\frac{c}{3}\frac{2\ddot{a}a+\dot{a}^2}{a^2}-8\pi p_M,
\label{equation003}
\end{eqnarray}
and
\begin{eqnarray}\label{equation102}
L&=&\frac{a}{\dot{a}}=\frac{1}{H},\ \ \lambda=\frac{4a^3c}{L^3},
\end{eqnarray}
From eqs.(\ref{equation002},\ref{equation003}), we can derive
\begin{eqnarray}\label{equation302}
\frac{\ddot{a}}{a}=-\frac{c}{3}\frac{\ddot{a}}{a}-\frac{4\pi}{3}(\rho_M+3p_M).
\end{eqnarray}
From eq.(\ref{equation002}), it is clear that $(1+\frac{c}{3})>0$.
Then eq.(\ref{equation302}) leads to $\frac{\ddot{a}}{a}<0$. So
there are no solutions of accelerating universe in this model.

\section{Conclusions}

In this paper, we propose a new HDE model with the action principle.
This is the first time that one can derive a HDE model from an
action. Furthermore, the causality problem has been solved
completely in this and the initial HDE models. By introducing two
fields $\lambda$ and $L$, all equations of motion become local and
the evolution of the universe is determined completely by present
initial conditions. So clearly the HDE models obey the law of
causality. Quite interestingly, the circular logic problem in HDE
can also be solved. One always criticizes the original HDE model
since the use of the future event horizon as a cut-off has assumed
the accelerating expansion of the universe. Then how can one use an
assumption based on the accelerating expansion to explain the
accelerating expansion? In fact, we do not need to assume the future
event horizon as the cut-off but only a local equation
$\dot{L}=-\frac{1}{a}$. Then the equations of motion magically force
the cut-off to be exactly the future event horizon. This is also a
support of the causality in HDE models. As we have shown, it is not
that the present evolution of the universe depends on the future
event horizon but that the future event horizon is determined by the
present cut-off through equations of motion.

It is interesting to note that this new model is very similar to the
initial one. For example, in both models, only the future event
horizon rather than the particle or Hubble horizons can drive the
accelerating expansion of the universe. Beside, both models have a
pure de Sitter solution and behaves as phantom below the critical
parameter( $c=6$ in the new model and $c^2=1$ in the initial one).
The main difference between the two models is that the new one
predicts the existence of dark radiation. We shall study this
interesting problem in the following works.

\section*{Acknowledgements}

We are grateful to X. D. Li, S. Wang and Z. H. Zhang for useful
discussions. M. Li would thank A. Zee for his early helpful
collaboration on the action principle of HDE. R. X. Miao would like
to thank X. D. Li for his valuable discussions. This research was
supported by a NSFC grant No.10535060/A050207, a NSFC grant
No.10975172, a NSFC group grant No.10821504.

\end{document}